\documentclass{article}
\usepackage[utf8]{inputenc}
\usepackage{graphicx}
\usepackage[section]{placeins}
\usepackage{systeme}
\graphicspath{{images/}}
\title{Probing Short Distance Gravity using Temporal Lensing}
\date{}
\author{Mir Faizal$^{1,2,3}$, Hrishikesh Patel$^4$
\\\\
\textit{\small $^1$Department of Physics and Astronomy,
University of Lethbridge,} \\ \textit{\small Lethbridge, AB T1K 3M4, Canada}
\\
\textit{\small $^2$Irving K. Barber School of Arts and Sciences, University of British Columbia}\\
\textit{\small Okanagan Campus, Kelowna, V1V1V7, Canada}
\\
\textit{\small $^3$Canadian Quantum Research Center, 204-3002, 32 Ave,}
\\
\textit{\small Vernon, BC, V1T 2L7, Canada}
\\
\textit{\small $^4$Department of Physics and Astronomy, University of British Columbia,} \\ \textit{\small 6224  Agricultural Road,
Vancouver, V6T 1Z1, Canada}
}
\begin{document}

\maketitle
\begin{abstract}
    \noindent It is known that probing gravity in the submillimeter-micrometer range is difficult due to  the  relative weakness of the gravitational force. We intend to overcome this challenge by using extreme temporal precision to monitor transient events in a gravitational field. We propose a compressed ultrafast photography system called T-CUP to serve this purpose. We show that the T-CUP's precision of 10 trillion frames per second can allow us to better resolve gravity at short distances. We also show the feasibility of the setup in measuring Yukawa and power-law corrections to gravity which have substantial theoretical motivation.
\end{abstract}
\section{Introduction}
General Relativity (GR) has been tested at large distances, and it has been realized that we need a more comprehensive model of gravity that aligns with our measurements of gravity at long distances and short distances. Several advances have been made in trying to understand gravity at long distances since the advent of multi-messenger astronomy \cite{multi1, multi2}. However, it is possible to relate the large distance corrections to the short distance corrections. In fact,  several theories like  MOND  \cite{MOND,MOND2} have been proposed to explain the large distance corrections to gravitational dynamics due to small accelerations. Furthermore, these theories can be tested at short distances, where such small accelerations can be produced experimentally  \cite{MOND4, MOND8}. Furthermore,  because  GR can be approximated by Newtonian gravity at short distances, 
and such short distance modifications in Newtonian gravity can indicate a  modification in  GR. Hence, it is crucial to test Newtonian gravity in the submillimeter-micrometer range. 
\par
Short distance modifications of gravity, at submillimeter scale are predicted from  brane world models with extra dimensions \cite{bran}-\cite{bran4}. It has been demonstrated that models with large extra dimensions can have Yukawa-like corrections at large distances \cite{yaka, yaka2}. Such Yukawa-like corrections have also been predicted in alternative theories of gravity like MOG \cite{MOG, MOG2}, $f(R)$ gravity \cite{mod1} and some generalizations of MOND \cite{mod2, mod4}. Such corrections can also occur at large distance galactic scales \cite{mod12, mod14} and it is expected that in theories like MOND, such corrections occur at very small accelerations, which can either be produced in short distances experiments, or at galactic distances.
Furthermore, in Randall-Sundrum models, the correction to the gravitational potential  is a power-law correction \cite{power, power2}. It is also possible to motivate the power law correction from GUP deformation of  the  entropic force \cite{gup, gup2}. Incidentally, there are several such theoretical models which can motivate such corrections to gravity and hence it is important to test such short distance modifications of the Newtons gravity. Such a test can at least potentially falsify or verify a class of theoretical models. 
\par 
Majority of the experiments proposed to test Newtons law at  short distances are based on the original  
Henry Cavendish's torsion balance experiment, which  was one of the first developments in testing gravity at short distances. However, since Cavendish's experiment, progress in probing gravity has been primarily made by improving the Cavendish experiment  \cite{submmgrav3}-\cite{submmgrav2} by instrumental modifications and slight design modifications. It has been known that probing gravity at smaller scales becomes increasingly difficult due to the weak nature of gravitational force. This is one of the reasons why probing gravity in the submillimeter-micrometer range is challenging using such conventional experimental setups. 
\par
Inspired by this challenge, we propose to use an optical setup with extreme temporal precision to probe gravity at short distances. It may be noted that the foundations of temporal imaging are based on the duality between paraxial diffraction and narrow band dispersion with quadratic phase modulation, as a result of this duality, temporal imaging is possible and a "time lens" is created \cite{TempImag}. This time lens enables temporal focusing much like a regular lens enables spatial focusing. The temporal precision of such a lens depends on the rate of emission of laser pulses. Due to massive advances in pulse compression and chirping mechanisms, it has been possible to  create ultra-short laser pulses \cite{pulsecomp1}-\cite{pulsecomp4}. These  advances in laser technology combined with the advances in streak camera technologies \cite{StreakCam}-\cite{StreakCam2} have enabled the creation of ultrafast photography systems \cite{CUP1}-\cite{CUP4}. These systems are capable of imaging repetitive or single shot events with picosecond and femtosecond accuracies. Due to these recent advances in the field of compressed ultrafast photography, a novel technology called T-CUP has been developed \cite{T-CUP}. T-CUP uses state-of-the-practice compressed sensing algorithms and streak camera technologies to capture transient events at ten trillion frames per second (10 Tfps). These advances have found extraordinary applications in the field of biology, chemistry and many other fields. In this paper, we propose to extend this technology to probe and study gravity at distances that are shorter than have ever been probed using the conventional methods \cite{submmgrav3}-\cite{submmgrav2}.

\par
Furthermore, to make our case more concrete, we also present numerical calculations for deviations predicted by Yukawa \cite{yaka, yaka2} and power-law  \cite{power, power2} modifications to Newtonian gravity. We further discuss some of the possible design challenges and ways to overcome them. Using the numerical data, we argue that the setup is indeed capable of reaching accuracies in the submillimeter-micrometer range. The paper is organized in the following sections: Sec II - Setup design and working, Sec III - Numerical Calculations for Yukawa modification, Sec IV - Numerical Calculations for Power-law modification, Sec V - Discussion and Summary.

\section{Setup}

\begin{figure}[h!]
$%
\begin{array}{cccc}
\includegraphics[width=130 mm]{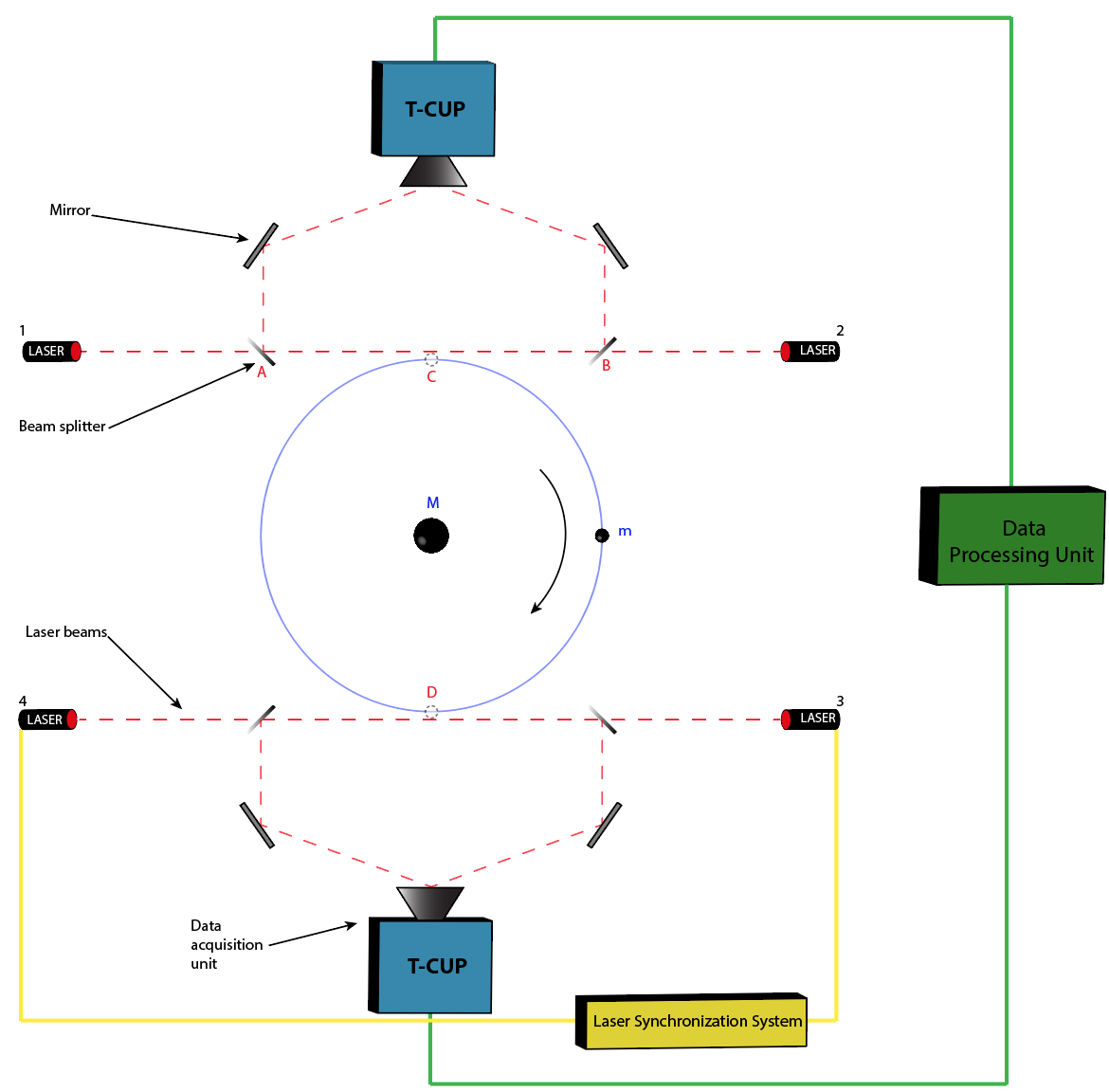}
\end{array}%
$%
\caption{Schematic model of the proposed setup}
\label{fig1}
\end{figure}

The proposed setup primarily consists of a particle performing uniform circular motion and a T-CUP system which monitors the dynamics of the particle. A particle of mass $m$ performs circular motion about the particle of mass $M$ $(M>>m)$ because  of gravitational force. If there are deviations from  Newtonian gravity, there will either be a shift in the time period of oscillation or the amplitude of oscillation. We propose the setup depicted in Fig.\ref{fig1} to measure these deviations. 

\begin{figure}[h!]
\begin{center}
$%
\begin{array}{cccc}
\includegraphics[width=120mm]{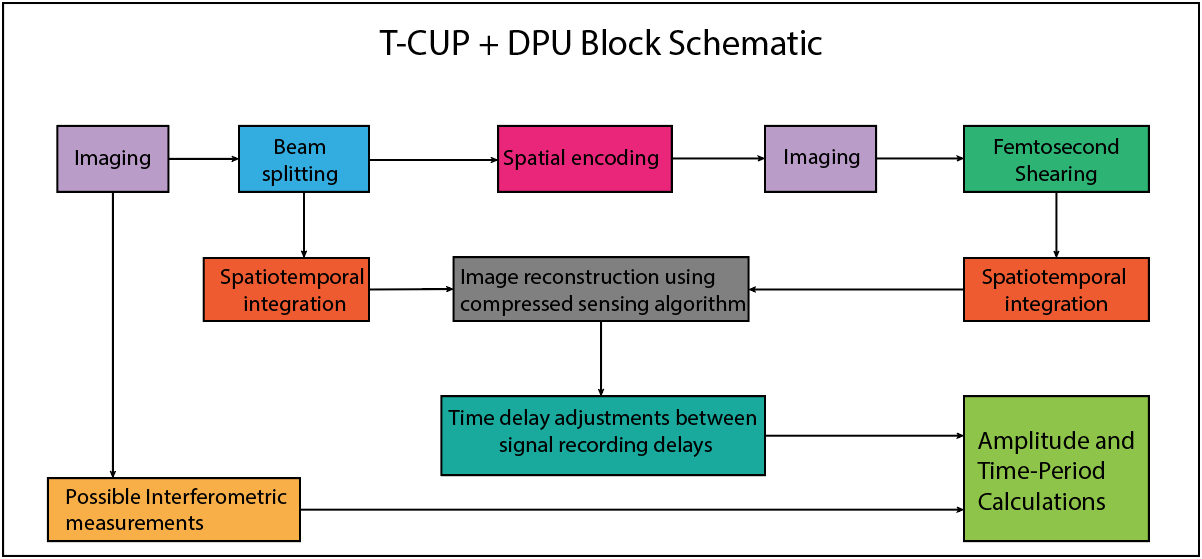}
\end{array}%
$%
\end{center}
\caption{Data Acquisition and Processing Schematic: \small This schematic is greatly inspired from the detailed diagram provided in \cite{T-CUP} with some relevant additions for the proposed setup}
\label{fig2}
\end{figure}

In Fig. \ref{fig1}, the upper half of the setup is completely identical to the lower half in terms of design and function. We use two such detection setups to accurately detect the time period and amplitude of the oscillation, furthermore, using two such detectors can help identify any directional systematic errors. The two pairs of lasers ($1$ and $2$) and ($3$ and $4$) emit laser beams in the opposite direction such that these beams are tangential to the circular path followed by the particle. When the particle is not at $C$, laser beam ($1$) transmits through beam splitter $A$ with half the intensity (the other half is reflected, which is unimportant for this setup), beam ($1$) then travels to beam splitter $B$ through point $C$. At $B$, half of the beam is reflected (transmitted part is unimportant) towards a mirror which redirects the optical signal into the T-CUP. Similarly, beam ($2$) transmits through $B$ and reflects through $A$, again hitting the T-CUP system in the end. Now, when the particle is at $C$, beam ($1$) transmits through $A$ and gets obstructed at $C$, similarly beam ($2$) transmits through $B$ and gets obstructed at $C$, thus no beam enters T-CUP. This is how a signal is registered. Once the signal is recorded into the data acquisition unit (DAU), the data is sent to the data processing unit (DPU). The DAU consists of a beam splitter, CCD (charge-coupled device) camera, Streak camera and a DMD (digital micromirror device). These perform the function of spatiotemporal integration and image encoding onto the DMD. This encoded image is then sent to the DPU for reconstruction using a compressed sensing algorithm. Once the images are constructed, time delay adjustments are made to account for the time difference between the events, i.e. particle at $C$ and signal recorded on T-CUP. Finally, computation of time period, amplitude and other parameters is done over many oscillation cycles to characterize the nature of gravitational force acting on the particle. Fig. \ref{fig2} gives a block representation of the complete data acquisition and processing procedure. 

\section{Yukawa-like corrections to Newtonian Gravity}
Below the micrometer scale, there are several insidious effects that might have the capability of modifying Newtonian gravity. Owing to such factors, many have proposed Yukawa-like corrections to Newtonian gravity. The gravitational potential is modified as \cite{yaka, yaka2}:

\begin{equation}
    \Phi_{Y}(r) = -\frac{\gamma}{r} \left( 1 + \alpha e^{-\frac{r}{\lambda}} \right)
\end{equation}
where $\gamma = GMm$, $\alpha$ is the strength parameter and $\lambda$ is the scale parameter. From classical mechanics, we know that the equation of motion for orbiting bodies is given by the following system of differential equations:
\begin{equation}
    \mu \ddot{r} = -\frac{\partial{\Phi}}{\partial{r}} + \frac{l^2}{\mu r^3}
\end{equation}
\begin{equation}
    \ddot{\phi} = \frac{l}{\mu r^2}
\end{equation}
where $\mu = \frac{Mm}{M+m}$ and $l$ is the total momentum. We assume that the total momentum of the system is constant. $\phi$ is the polar coordinate and $\Phi$ is any given gravitational potential. For Newtonian potential $\Phi = -\frac{\gamma}{r}$, we get:

\begin{equation}
    \ddot{r} = -\frac{\gamma}{\mu r^2} + \frac{l^2}{\mu^2 r^3}
\end{equation}

Similarly, for Yukawa corrections, we get the following differential equation:

\begin{equation}
\ddot{r} = -\frac{\gamma}{\mu r^2} + \frac{l^2}{\mu^2 r^3} - \frac{\gamma \alpha}{\mu r^2}e^{-\frac{r}{\lambda}} - \frac{\gamma \alpha}{\mu \lambda r}e^{-\frac{r}{\lambda}}
\end{equation}

\begin{figure}[h!]
\begin{center}
$%
\begin{array}{cccc}
\includegraphics[width=90mm]{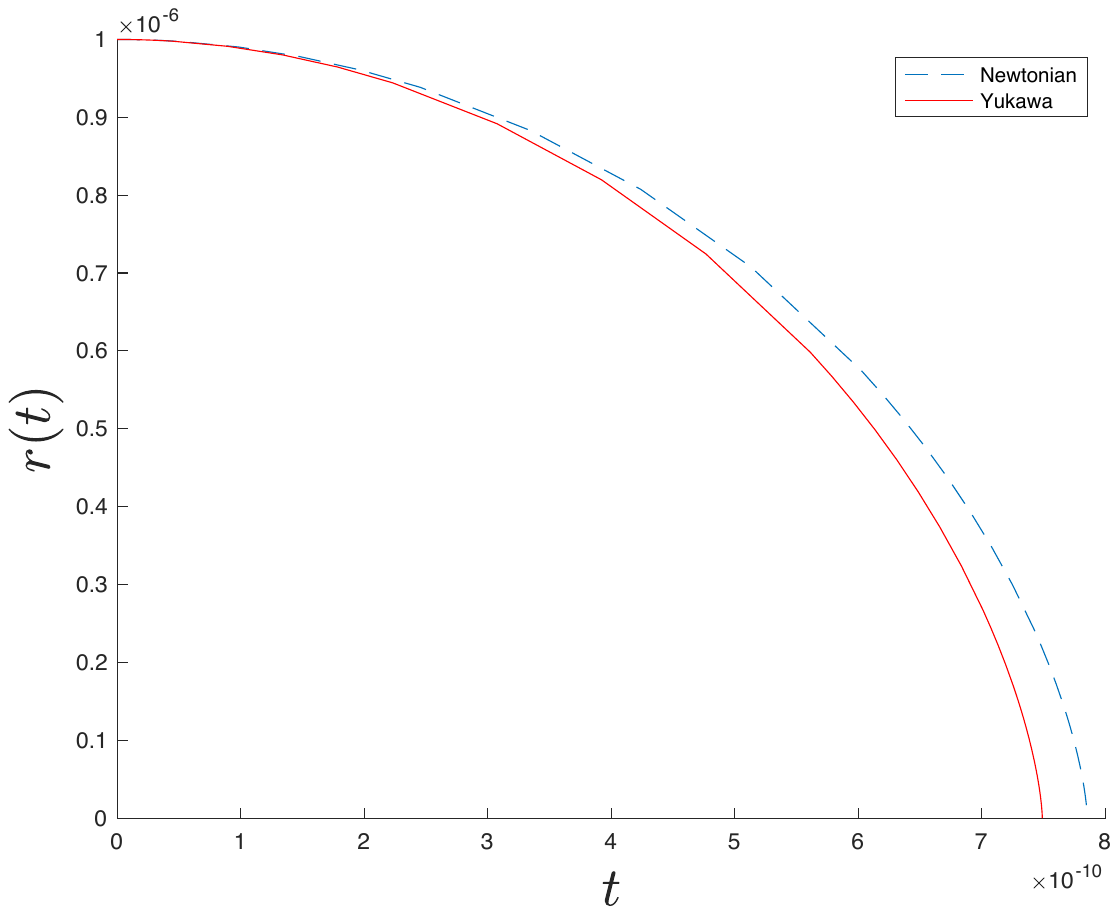}
\end{array}%
$%
\end{center}
\caption{For the free fall trajectories, we choose $\alpha = 1$, $\lambda = 10^{-5}$m, $\mu = 0.5$ kg and initial distance to be $10^{-6}$m.}
\label{fig3}
\end{figure}

The above equation cannot be solved analytically and it is very difficult to solve numerically due to exponential convergence. However, it can be solved for some special cases. To demonstrate this, we solve it for $l=0$, which is basically free fall as shown in Fig. \ref{fig3}, where the Newtonian trajectory has the familiar parabolic shape, whereas the Yukawa trajectory mimics the Newtonian trajectory, but with collision happening sooner. Although, these set of initial conditions are not consistent with the setup we proposed in the previous section, it must be noted that with some minor changes in the setup, we can measure deviations in free fall trajectories. This can be easily be observed from Fig. \ref{fig3}, where the temporal difference in collision times is of the order of $10^{-11}$s, which is measurable using the proposed setup, which reaches temporal sensitivity of $10^{-15}$s. One disadvantage of using the current setup of Yukawa corrections is that in order to measure deviations at the scale of $\lambda = 10^{-5}$m, the initial distance between the particles has to be about 1 micron. This can pose serious challenges especially in terms of setup construction and other systematic noises at small scales, these issues can be severely amplified when we try to make measurements below these scales. As we will see in the next section, this is not the case with power-law corrections, which are easily measurable (if they exist) due to their non-exponential convergence.

\section{Power Law corrections to Newtonian Gravity}
Several alternative theories have proposed power law corrections to Newtonian gravity and hence it is essential that we test the capability of the proposed setup in detecting any power-law deviations from Newtonian gravity at short distances. As we will see in this section, our setup is best suited for measuring such power-law corrections to Newtonian potential with sensitivities well in the micrometer range. 
The modification of the gravitational potential energy is as follows  \cite{power, power2}:

\begin{equation}
    \Phi_{PL}(r) = -\frac{\gamma}{r} \left( 1+\frac{k}{r} \right)
\end{equation}

\noindent where k is the length parameter. Deviations from Newtonian gravity are only significant (measurable) when $r \approx k$. Currently, the bound lies at $k = 52 \mu m$ \cite{submmgrav3} - \cite{submmgrav2}, which are set by the conventional torsion balance systems. Now, in order to solve Eq. 6, we follow a slightly different procedure then we did for Yukawa corrections, instead of finding $r(t)$, we try to find $r(\phi)$ by working in polar coordinates. For this, we convert all time derivatives into $\phi$ derivatives and make the basis transformation $u=\frac{1}{r}$. Hence, for the Newtonian potential, Eq. 4 transforms to:

\begin{equation}
    \frac{d^2u}{d \phi^2} = -u(\phi) + \frac{\mu \gamma}{l^2}
\end{equation}
 Solving the equation, we get the trajectory equation to be:
\begin{equation}
    r(\phi) = \frac{c}{1+\epsilon cos(\phi - \delta)}
\end{equation}
where $c = \frac{l^2}{\mu \gamma}$, $\epsilon$ controls the nature of the orbit (circular, elliptical, hyperbolic, etc.) and $\delta$ orients the orbit. This polar equation will be useful when we analyze circular trajectories. Furthermore, on repeating the calculations for $\Phi_{PL}$, we get the following equation:

\begin{equation}
\frac{d^2u}{d \phi^2} = -u \left( 1- \frac{2\mu \gamma k}{l^2}\right) + \frac{\mu \gamma}{l^2}
\end{equation}

\noindent Fortunately, unlike the Yukawa case, this equation can be solved analytically. To simplify the notation slightly, we define $a = \frac{2 \mu \gamma k}{l^2}$ and $b = \frac{\mu \gamma}{l^2}$. The solution of which is given by:

\begin{equation}
    r(\phi) = \left[ \frac{b}{1-a} + c_1 e^{\sqrt{a-1}\phi} + c_2 e^{-\sqrt{a-1}\phi} \right]^{-1}
\end{equation}
where $c_1$ and $c_2$ are set by initial conditions (much like $\epsilon$ and $\delta$). Moreover, $a = \frac{2k}{c}$ and $b = \frac{1}{c}$, which can change with the masses and the initial momentum (since we assume the momentum to be constant). These changes in $a$ and $b$ can change the nature of the trajectory of the particle. For example, choosing $c_1 = c_2 = 0$ gives us a circular orbit. To make our calculations more meaningful and consistent with the setup we proposed in the earlier sections of this work, we choose the parameters ($M,m,l,\delta$) that give us a circular orbit. Hence, we get:
\begin{eqnarray}
 c_1&=&\frac{a - \sqrt{a-1}}{2c(a-1)}e^{-\sqrt{a-1}\pi/2} \nonumber\\
 c_2&=&\frac{a + \sqrt{a-1}}{2c(a-1)}e^{\sqrt{a-1}\pi/2}
\end{eqnarray}

\begin{figure}[h!]
\begin{center}
$%
\begin{array}{cccc}
\includegraphics[width=100mm]{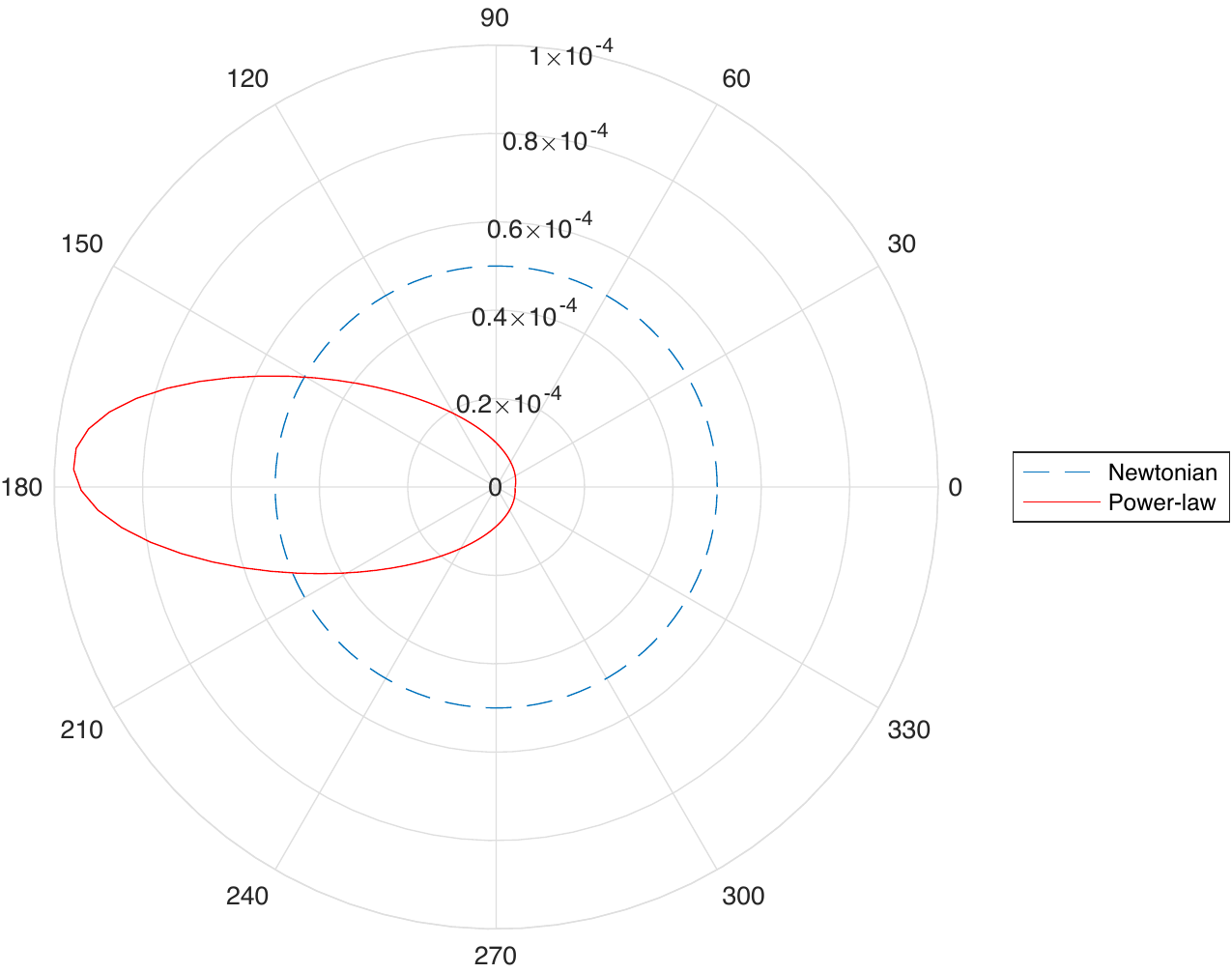}
\end{array}%
$%
\end{center}
\caption{As delineated in the figure, power-law modifications change the nature of the trajectory in a measurable way. The plot is exaggerated due to polar scaling. The maximum difference in the distance between the trajectories at a particular time is $<10^{-5}$ as portrayed in Fig. \ref{fig5}. Nevertheless, it demonstrates the tractability of possible deviations. Here, $k=1\mu m$, $c=10\mu m$, $\epsilon = 10^{-8}$ (almost circular).}
\label{fig4}
\end{figure}

Calculating trajectories for the Newtonian law and the power-law corrections numerically, we get the trajectories as displayed in  Fig. \ref{fig4}.

\begin{figure}[h!]
\begin{center}
$%
\begin{array}{cccc}
\includegraphics[width=100mm]{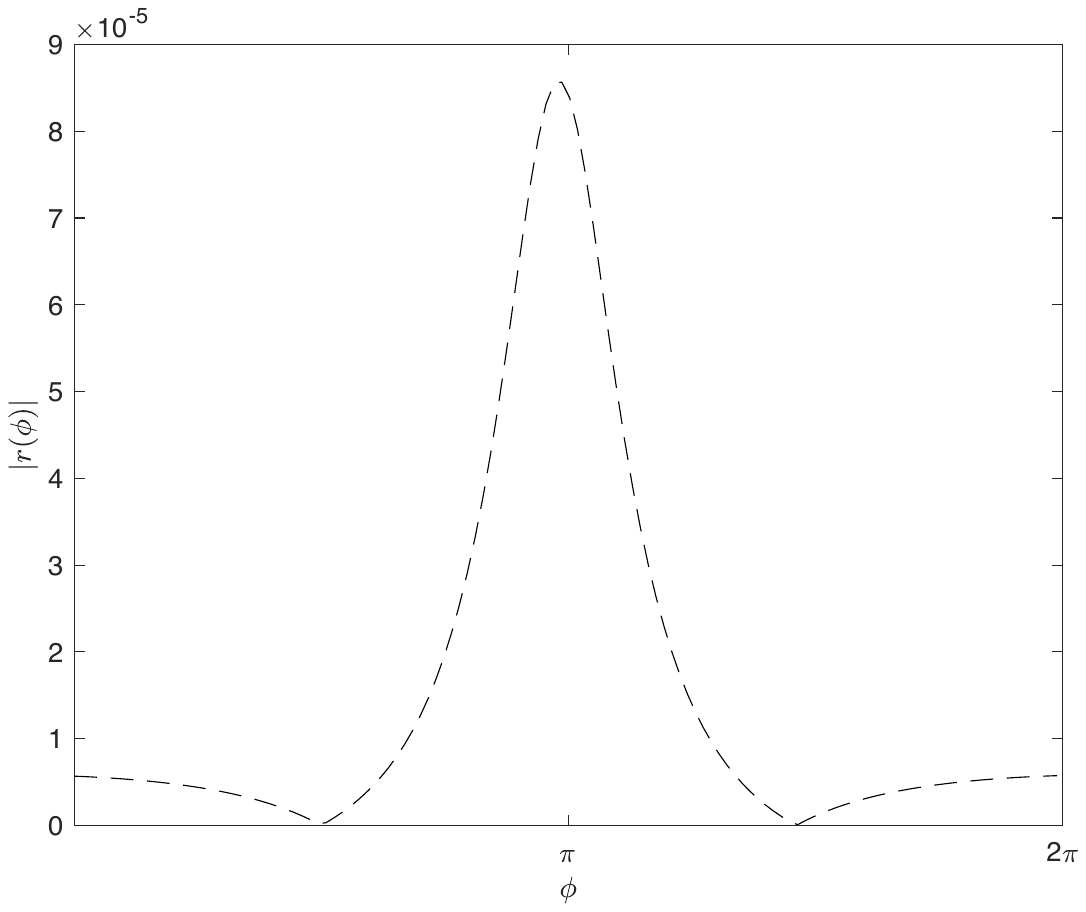}
\end{array}%
$%
\end{center}
\caption{The above plot portrays the difference in trajectories due to the power-law modification as a function of $\phi$. Based on the chosen initial conditions, the difference is maximum at $\pi$ radians}
\label{fig5}
\end{figure}

The portrayal in Fig. \ref{fig4} is exaggerated due to polar scaling, nevertheless, it is obvious that the power-law modifications deform the trajectory in a measurable manner. As shown in Fig. \ref{fig5}, the difference in trajectory due to the power-law modifications can go down in the micrometer range depending upon $\phi$. Since, the setup is completely optical, a spatial shift of 1 micrometer can be translated to a temporal shift of $10^{-14}$s, which is within the temporal sensitivity of the T-CUP system. 

This is a significant result. It reaffirms our initial claim about the usefulness of the T-CUP system. These ultrafast photography units are very recent and the only systems that can reach femtosecond sensitivities. Thus, practical applications of these systems for measuring gravity in the submillimeter-micrometer scale have never been studied or even speculated before. Additionally, this is still a 'work in progress' field and it has seen major advances in laser technologies, streak camera technologies and compressed sensing algorithms which provide further scope to improve the T-CUP systems to reach even deeper sensitivities. Furthermore, even though efficient miniaturization of the setup is in general a massive challenge, the numerical results suggest that in order to reach sensitivities in the micrometer range (i.e. $k \approx 10^{-6}$m), the orbital setup itself can be bigger than the scale itself (i.e. $c \approx 10^{-5}$m), which is realistic and can be especially advantageous if we seek to probe shorter distances in future. 

\section{Discussion \& Summary}

It may be noted that the setup we propose is a working model. However, modifications to this setup could be made in order to make it more suitable for testing various different models of gravity. A clear advantage of the system is that it overcomes the challenge of directly probing short distances, instead it equivalently probes short time-scales. Furthermore, as it is a passive setup, we do not introduce a probe in a way that would disturb the system. To illustrate this, when the particle is at $C$, the two sets of lasers apply an equal and opposite force on the particle. As a result, the particle experiences no net force except gravity during the complete trajectory. Additionally, T-CUP has been designed in a receive only fashion to record transient events, thus it is necessary that we create a sudden burst or a sudden void in the energy signal in order to detect the particle accurately without disturbing it. Hence, Fig.\ref{fig1} is a good arrangement because it complements the T-CUP system and enables us to attain deeper gravitational sensitivity. The simplicity of the proposed setup also allows efficient control of the system at smaller scales compared to the conventional setups. As the system has less moving parts, miniaturization and modification of the setup can easily be done in order to probe any particular model of non-Newtonian gravity at short distances. Moreover, it is known that optical systems can achieve higher precision compared to mechanical systems like torsion balance, these systems also do not face the disadvantage of dissipative forces like friction. 
\par
As an additional comment, some readers might have noticed that the setup design has a striking similarity to an interferometer. In fact, with some additional modifications, the setup can also be used as an interferometer. This would allow the system to use extreme temporal and spatial precision to probe gravity at very short distances. The interferometer cannot however be used as a general purpose gravitational setup but rather as an instrument to test for specific asymmetrical corrections to Newtonian gravity. Thus, we do not discuss it here. 
\par
In order to prevent significant impact of earth's gravity and other sources of noise on the system, we propose to conduct such an experiment in outer space. We also intend to employ some traditional methods to shield electromagnetic forces using methods similar to the ones suggested in \cite{submmgrav} and \cite{submmgrav2}. We can also calibrate the system to account for the known tidal forces exerted by nearby celestial objects. Gravitational calibration methods similar to the ones suggested in \cite{submmgrav} can also be used for this purpose. Moreover, the shape of the trajectory can also be altered based on convenience and depending upon the gravitational model under discussion. In fact, according to Kepler's law, the general shape of the orbit is elliptical, and it seems likely that the deviations could be significant with open trajectories. However, closed trajectories allow acquiring multiple iterations of data, which can give more precise results in the end. Moreover, due to the additional symmetry provided by spherical trajectories, data can be taken at several points and can be easily analyzed and rectified for due to anisotropic corrections that may arise because of several factors (EM field, weak gravitational effects due nearby celestial objects). There is a possibility of doing the experiment on the Earth with suitable cryogenic gyroscopes that were designed for the Gravity Probe B (GPB) experiment \cite{GPB}. Although, it must be noted that GPB itself was done in outer space, and so we think it is advisable to do such an experiment in the space, especially since the basic technology/infrastructure already exists.
\par
We understand that there still might be plethora of minor experimental barriers in performing such radically new experiment and it is not possible to foresee or discuss all those experimental challenges in this preliminary work. Nevertheless, the experiment has critical advantages over conventional methods and it seems plausible that we can probe gravity at scales below $10 \mu m$ using such optical setups. Probing gravity at this scale could allow us to understand gravity at a much deeper scale. Some of the direct consequences of this experiment would be elimination of various gravitational models, more stringent bounds on the large extra higher dimensions   \cite{yaka,yaka2}, or Randall-Sundrum model \cite{power, power2}. It would also be interesting to propose similar tests for  MOND \cite{MOND,MOND2} and other modified gravity theories \cite{mod1,mod2,mod4,mod12,mod14}. Furthermore, such measurements could in turn help us better understand many cosmological puzzles like dark matter and dark energy. It might also be possible to consider corrections which break the  of isotropy of space, and such corrections could then be used to understand the cosmic anisotropies in the Planck data \cite{data, data2}.

\section{Author statement}
H.P. and M.F. jointly conceived the proposed setup and the underlying ideas for the model. H.P. did the primary drafting of the manuscript along with diagram creation, and the relevant numerical analysis. M.F. provided timely suggestions and participated in the final stages of the project like editing and reviewing of the manuscript. Both the authors have agreed upon the results and ideas presented in the manuscript. None of the authors have any competing interests, financial or otherwise.
{}
\end{document}